# Understanding communications in medical emergency situations

Lyuba Mancheva
GIPSA-Lab / LPNC / Univ. of Grenoble Alps /
Laboratoire d'Etude des Mécanismes cognitifs,
Univ. of Lyon 2
Lyuba.Mancheva@gipsa-lab.grenoble-inp.fr

Julie Dugdale
Univ. of Grenoble Alps, France /
Grenoble Informatics Lab (LIG) /
Agder Univ. Norway
Julie.Dugdale@imag.fr

**Abstract**

*Good communication is essential within teams dealing with emergency situations. In this paper we look at communications within a resuscitation team performing cardio-pulmonary resuscitation. Communication underpins efficient collaboration, joint coordination of work, and helps to construct a mutual awareness of the situation. Poor communication wastes valuable time and can ultimately lead to life-threatening mistakes. Although training sessions frequently focus on medical knowledge and procedures, soft skills, such as communication receive less attention. This paper analyses communication problems in the case of CPR and proposes an architecture that merges a situation awareness model and the belief-desire-intention (BDI) approach in multi-agent systems. The architecture forms the basis of an agent-based simulator used to assess communication protocols in CPR teams.*

## 1. Introduction

The importance of good communication is one, if not the most, critical issues and fundamental activities in emergency management. Good communication underpins the supply of information, effective collaboration and the coordination of tasks, not to mention its role in supporting leadership and direction [20] [30]. Communication is a social interaction and can be viewed as a dynamic complex system where entities, e.g. humans, interact via a common code (language, gestures, sounds, facial expressions, etc.). These interactions can be studied both at the micro level in how an individual processes information, and at the macro level, in how communication dynamics emerge in a group of individuals and how they contribute to a shared situation awareness.

Various research domains (e.g. linguistics, philosophy of language, computer science, psychology, systems science, and anthropology, to name but a few) analyse and aim to understand communication, its problems and its consequences, within a group of entities in a dynamic situation. Such studies are commonly applied to real life problems in aviation, emergency response, and military actions, etc. Although the notion of everyone using a common communication code, problems frequently occur because the entities involved in these dynamic processes interpret information differently depending on their mental attitudes and abilities. This adversely affects knowledge and information sharing.

This work aims to understand how communication is linked to the performance of medical emergency teams and what possible solutions could be used for its improvement. The final goal is to develop an agent based computer simulator that will allow trainers of resuscitation teams to experiment with different communication protocols. Agent-based modelling (ABM) has seen an enormous growth in popularity in recent years. ABM is concerned with developing a computational model for *simulating* the actions and interactions of a group of autonomous individuals, for example humans. One of the main advantages of this approach is that it gives us the opportunity to explore and experiment with social situations that we might not be able to do in real life. A simulation model can be set-up and then executed many times, varying the conditions in which it runs and exploring the effects of different parameters. These advantages, coupled with the difficulty in modelling social situations using a top-down mathematical approach, make ABM a suitable choice for our particular problem.

The methodology adopted for our overall work is strongly iterative and is composed of 5 steps:
1. Conduct field studies to analyse communications within resuscitation teams.
2. Analyse communications from the field studies
3. Develop a formal model of communications and a cognitive model of entities within the simulator

4. Implement the model in a computational simulation platform
5. Experimentation

Overlaying each of these steps is *validation;* validation of field studies data and the analysis by cross reference to other data and theory; validation of the formal models with alternative field studies data sets and with experts, validation of simulator outputs by sensitivity analysis, expert focus groups and with cross referencing to field results.

Before concentrating on the description of the cognitive model, which is the focus of this paper, we briefly summarise our work on the first two steps and the implementation of the model.

Concerning the field studies, since we cannot collect data on communications during real-life CPR procedures for ethical reasons, we have been collaborating with the local hospital training centre for the region to collect data on their simulation exercises. We analysed the communications exchanged during the training simulations and found that they heavily affected the success of the team [21, 22].

To mode the communications we categorized each communicative item, with our extended version of the FIPA-ACL (Agent Communication Language) [11]. In collaboration with the physicians, we then identified trends of what constituted good and bad communication acts in resuscitation teams. As an example, we found that that the total number of messages exchanged was significantly lower in teams that performed well compared to teams that performed badly. Also, badly performing teams do not wait for a fact to be known to be true before performing an action [22].

Following our iterative approach a basic cognitive model of the agents was developed in order to implement a first version of the simulator using the RePast S platform [28]. The model has been subsequently developed and it is this model and its interplay with communication that is the subject of this paper.

Several cognitive architectures exist to describe the structural properties of a cognitive system for computational implementation and two popular ones are SOAR and ACT-R [29] [5]. However, the architectures suffer from several problems, either they are not suited specifically to multi-agent social simulation, or they are based on underlying knowledge formalism (such as procedural rules) that does not adequately capture the complexity of human behaviour, or the implementation toolkits and frameworks that are available cannot easily cope with these architectures [1] [40].

Our approach is to use a flexible Belief-Desire-Intention (BDI) architecture [14]. BDI is based on folk psychology and designs agents in terms of their underlying beliefs about the situation, their desires i.e. things that they wish to do in the future, and their intentions i.e. their chosen actions that will allow them to achieve their goals [27].

The paper is organised as follows: section 2 describes the process of cardio-pulmonary resuscitation CPR and looks at the main communication issues experienced by the resuscitation team. One of the main cornerstones of successful decision-making and teamwork is situation awareness and mutual situation awareness and it is in section 3 that we discuss these notions. We then show how Endsley's situation awareness model [8] [9] has been adapted to incorporate a multi-agent system BDI architecture. Finally we conclude this paper with a summary of contributions and a discussion on future work.

## 2. Cardio-pulmonary resuscitation (CPR)

When the human heart fails to contract effectively and there is a sudden interruption in blood circulation, a victim enters the phase of cardio-pulmonary arrest. Due to the lack of oxygen to the brain the patient becomes unconscious and breathing is abnormal or ceases altogether. Without the correct treatment, irreversible brain lesions may develop and death often follows.

Bystanders or paramedics may provide on-site Basic Life Support (BLS) for cardio-pulmonary resuscitation. This involves giving chest compressions, providing artificial respiration and, if possible, restoring the heart rhythm by using an automated and portable defibrillator machine.

BLS is maintained until Advanced Cardiac Life Support (ACLS) can be provided in a hospital or on-site by a qualified medical team. If a victim is already in hospital when the cardio-pulmonary arrest occurs, ACLS is started immediately. ACLS differs from BLS in that additional procedures, such as tracheal intubation, more sophisticated cardiac monitoring, and intravenous cannulation are usually undertaken.

Resuscitation teams follow a specific algorithm, usually expressed as a flow-chart, to perform CPR. Although there are small differences in the algorithm to take into account CPR in children for example, or because of national guidelines, the following basic iterative steps are always performed: after assessing responsiveness, looking for signs of life and checking for obstructions in the airway, start CPR with chest compressions and ventilations with a ratio of 30:2 (for adults); if the rhythm is shockable stop compressions and ventilations and perform shock; resume compressions and ventilations for 2 minutes; assess rhythm and perform shock, repeat. Medicines are usually administered intravenously during CPR; Adrenaline after the 3$^{rd}$ shock and Amiodarone after the 4$^{th}$ shock if there is still no pulse. CPR is stopped after several iterations if there is no sign of life, or if the leader decides that further CPR is futile.

Some medical facilities now have dedicated rapid response resuscitation teams that are composed of healthcare clinicians; typically they include a physician with an assistant, a critical care nurse, clinical nurse, and a respiratory therapist. The formation of such teams has increased survival rates [18].

Training resuscitation teams is of paramount importance since early intervention, speed, and knowledge of the appropriate actions to perform are essential for increasing a victim's chances of survival. Although survival rates are somewhere within the range of 13 to 59%, some studies have reported rates as low as 4% [35]. A major contributing factor to the low rates is a lack of basic resuscitation skills in doctors and nurses [16]. This has prompted extensive training programmes in hospitals and more frequent reviews to procedural guidelines, such as those given by the American Heart Association and the European Resuscitation Council. These measures are taking effect. A 2011 study showed that formal training of the CPR team improves survival rates and survival to hospital discharge, noting a difference between pre BLS and post ACLS training from 18.3% to 28.3% in the return of spontaneous circulation in victims [18]. Despite this increase and given our advanced medical knowledge of cardiology, survival rates still remain low with death being the most likely outcome for the majority of victims.

Traditionally, increasing survival rates has been sought by focusing on education and training of resuscitation in order to ensure that the right tasks are performed in the correct order and in good time. However, in recent years, the emphasis on training has shifted from instilling medical knowledge to the application of non-technical skills (NTSs) or soft skills [4], [10], [38].

Many studies have shown that non-medical skills and in particular communication are essential for the efficient working of the resuscitation teams [3] [39], [24], [20]. One study reported that 43% of medical errors in surgery are due to communication failures or inadequate communication [13]. Looking specifically at teamwork in healthcare, another study reviewed 277 articles published between 1950 and 2007 and summarized the non-medical skills needed for healthcare teams [23]. Figure 1 shows the conclusions of the study, citing leadership, collaboration, communication, coordination, shared mental model and leadership as being paramount in increasing quality and safety of patient care.

| Aspects of teamwork | Examples of safety-relevant characteristics |
|---|---|
| Quality of collaboration | Mutual respect<br>Trust |
| Shared mental models | Strength of shared goals<br>Shared perception of a situation<br>Shared understanding of team structure, team task, team roles, etc. |
| Coordination | Adaptive coordination (e.g. dynamic task allocation when new members join the team; shift between explicit and implicit forms of coordination; increased information exchange and planning in critical situations) |
| Communication | Openness of communication<br>Quality of communication (e.g. shared frames of reference)<br>Specific communication practices (e.g. team briefing) |
| Leadership | Leadership style (value contributions from staff, encourage participation in decision-making, etc.)<br>Adaptive leadership behavior (e.g. increased explicit leadership behavior in critical situations) |

Overview of aspects of teamwork relevant to the quality and safety of patient care in dynamical domains of healthcare.

**Figure 1: Results of the retrospective analysis concerning teamwork and patient safety in the healthcare domain [23]**

Reflecting a little deeper on this figure we notice that almost every aspect of good teamwork relies upon communication effectiveness. Communication facilitates collaboration through the exchange of the information that is necessary in order to work together as a team. Communication allows team members to articulate their roles and negotiate tasks with respect to the joint coordination, for example in the redistribution of roles (role-shifting). Communication is needed in leadership to make explicit how the team will be practically organised. Finally, communication supports the sharing of mental models by making public an individual's perception of the situation. A mental model describes a person's thought process and is a representation of elements in the surrounding world. Here we can see a close affiliation between mental models and the BDI architecture that used in agent based systems. Like a mental model the BDI architecture is concerned with human practical reasoning and representing mental attitudes.

Mental models are a prerequisite for achieving situation awareness (SA). SA is commonly defined as being aware of what is happening in the environment, in order to understand how information, events, and one's own actions will impact goals and objectives, both immediately and in the near future. SA is an important concept in decision-making in complex, dynamic environments such as medical emergency situations.

## 3. Situation awareness (SA) and mutual situation awareness (MSA)

A seminal and widely used model of SA is was proposed by Endsley that breaks SA down into 3 levels [8] (figure 2):
1. Perception of the elements in the current situation (level 1). This is a crucial level that concerns recognising current conditions and elements in the environment, without it subsequent levels cannot function.
2. Comprehension of the current situation (level 2). At this level individuals are concerned with the process of understanding what has been perceived; it involves combining, interpreting and storing information.
3. Projection of future status. This is the ability to project what will happen in the future and to plan for future states.

The perception of time and the temporal dynamics of events play a very important role in the construction of SA.

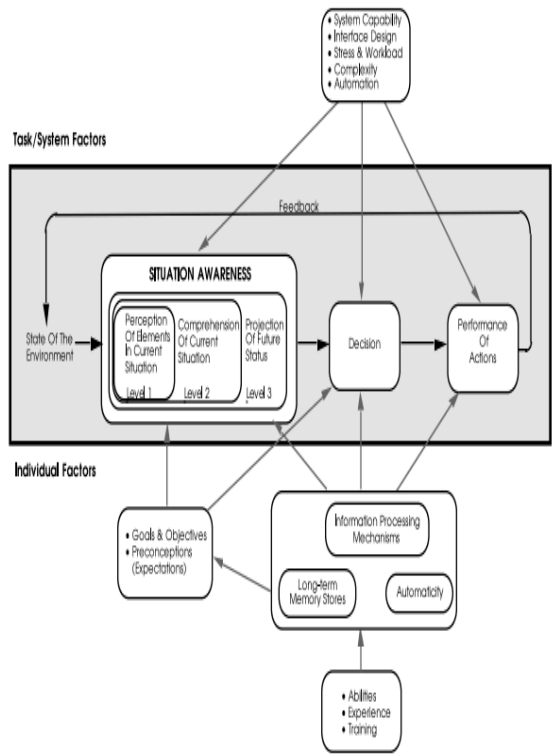

**Figure 2: Situation awareness model [8]**

Despite the intuitive simplicity and explanatory power of these levels, there are some problems with Endsley's model. The first is that the cognitive processes of perceiving, understanding and projection expressed in the 3 levels are only described in general terms and are not operationalized [12]. The second problem is that the levels are treated as sequential. In the real world comprehension starts with the first perceptual input and does not wait until all percepts are received. [19]. Finally, situation awareness depends on the notion of shared cognition [6]. In essence, Endsley's approach does not model the shared mental model of the situation of many people; instead it describes the internal model of the situation of a person.

Concerning the issues of a lack of operationalization and explicit level sequence we will show in section 5 how a BDI architecture can overcome these problems. The problem of shared mental models necessitates addressing the issue of mutual situation awareness (MSA).

Mutual situation awareness employs a double feedback process: monitoring events and the actions of other people in the environment, and, at the same time making visible one's own actions that are relevant to other people [32]. It is not necessary that team members are aware of everything at all times, but that the right information is sent to the right person at the right time via coordination [18], [9]. However receiving more data does not necessarily mean having more information [9]. Because of MSA, team members share a common representation of the environment, the team structure and the team's tasks. Team members should exchange information and knowledge to build and maintain a shared mental model of the situation; this exchange is possible through communication.

[36] found that the perception of physicians and other team member is not the same, noting that physicians often tend to overestimate teamwork and communication. This could indicate that they have built an incorrect picture of the situation because they do not effectively communicate with their teams.

## 4. Communications and MSA

Many studies identify the solid link between communication, and SA and MSA that strongly affects a team's performance [15].

Communication failures maybe classified along several dimensions: the content (messages are unclear or unstructured) [7] the entities themselves (members of the team do not always share the same communication code [3], and the intended purpose of the communication (members do not always correctly interpret the message). Other reasons for communication failures arise from an excessively noisy environment, stress, and the lack of accurate communication strategies in a particular critical situation, etc. [26].

The Agency for Healthcare Research and Quality (AHRQ) in the United States recommends several information exchange standards for effective communication that will increase MSA [2] (Table 1)

**Table 1: Techniques for effective communication**

| Information exchange standards | Description |
|---|---|
| SBAR | A technique of structuring critical spoken information. SBAR refers to the situation (the current state of the patient), Background (the patient's clinical context), Assessment (what is wrong with the patient), and Recommendation (suggestions for treatment). The following SBAR was defined by the resuscitation team trainer and used as a scenario in the simulation exercises that were studied as part of this work: Situation – an unconscious man has been given BLS and cardiac massage for 3 to 4 minutes by a member of the public. Background – the man, who is about 50 years old, complained of chest pain before collapsing. Assessment – cardiac arrest. Recommendation – start ACLS. |
| Call-out | A tactic used to distribute critical information *during* an urgent event. This allows other members to anticipate their future actions. |
| Handoff | Occurs when patient care is transferred to other health-care professionals. The completeness and certainty of the information transmitted during handoff is of utmost importance, making handoffs open to communication breakdowns. |
| Checkback | Closed communication loop, used to verify and validate the information transmitted. This strategy allows the sender of the message to check if the message is received and for the receiver to confirm acceptance of the message and its content. |

These techniques allow members of resuscitation teams to ensure more efficient communications by imposing a precise sequence or a specified content of the communication. It is the effectiveness of these techniques that we would like to assess with our simulator. Specifically, to answer questions such as: how much does MSA increase or decrease as a result of using these techniques? What is the overhead, for example, in terms of time of using these techniques? Rather than preventing errors, could they introduce new ones? Etc.

To highlight the problem with bad communications the following figure gives an example of a good and bad communication sequence. The example is an excerpt from our field studies data. The left-most column shows bad communication. Note that the sequence is to be read top to bottom. The centre column shows a good communication sequence, top to bottom, of the same set of actions during CPR. Note that the terms 'good' and 'bad' were expressly chosen by the resuscitation team trainer (a physician); this indicates the importance placed on communication by the trainer. Also shown in the rightmost column is an example of how we have translated the good communications into a logic-based message that forms part of our model [21] [22]. It is possible with our formalism also to represent bad communications.

| Bad communication sequence | Good communication sequence | Logical representation of the good messages |
|---|---|---|
| *Prepare what you need to shock if he is fibrillating* (physician, nurse) | *It is a shockable rhythm, he is fibrillating* (physician, nurse) | **inform** (physician, all, fibrillation (patient)) |
| *Massages again while we...* (physician, critical care nurse) | *I want you to administer the shock* (physician, nurse) | **request** (physician, critical care nurse, shock (critical care nurse, patient)) |
| *We are going to recharge* (physician, critical care nurse) | *Are you ready to shock?* (physician, nurse) | **query-if** (physician, nurse, ready (nurse, shock (nurse, patient))) |
| *Massage, massage* (physician, critical care nurse) | *Yes* (nurse, physician) | **confirm** (nurse, physician, ready (nurse, shock (nurse, patient))) |
| *Everyone move away* (physician, all) | *Stop massaging* (physician, critical care | **request** (physician, critical care nurse, stopChestCompre |

|  | nurse) | ssions (critical_care_nurse, patient) |
|---|---|---|
| *Be careful I'm shocking* (physician, all) | *We are going to shock* (physician, all) | **inform** (physician, all, shock (nurse, patient)) |
| *Let's massage another minute* (physician, all) | *Step back* (physician, all) | **request** (physician, all, move (all)) |

**Figure 3: An excerpt from our field studies data showing both bad (left-most column) and good (central column) communicative interactions. The rightmost column shows the representation of the good communications into a logic-based message that forms part of our model.**

In the first sequence the messages are unclear, the physician did not finish the sentence, and the identity of the person who will perform the shock is unclear (either the critical care nurse or the nurse). Consequently, the critical care nurse thinks that the physician asked her to perform the shock, so she stops the massage in order to prepare the shock. Therefore no one is performing the massage any longer and the physician has to repeat himself to target the right person). This is actually a serious problem since stopping massage leads increases 'No Flow' time, i.e. the time when there is no blood flow. In the second sequence the messages are well structured, the doctor asks the nurse if she is ready to make the shock, he awaits her confirmation. Also in good communications, the physician maintains the MSA using informative messages to everybody (e.g. "*It is a shockable rhythm; We are going to shock*")

In addition to the effectiveness of the recommendation techniques mentioned in table 1, we would also like to assess in a more efficient way through simulation, exactly where the main problems arise in bad communication sequences.

## 5. Integrating SA and MSA into an agent based BDI MODEL

In order for our work on understanding the role of communication and its problems to be put into practice, we have developed a cognitive architecture for our agents that forms the basis of our simulator (Figure3) shows the cognitive architecture of our agents in our system. In the architecture we have incorporated the classic BDI architecture with Endsley's SA model. The orange blocks represent the different levels in the situational awareness model; these in turn contain the underlying BDI elements.

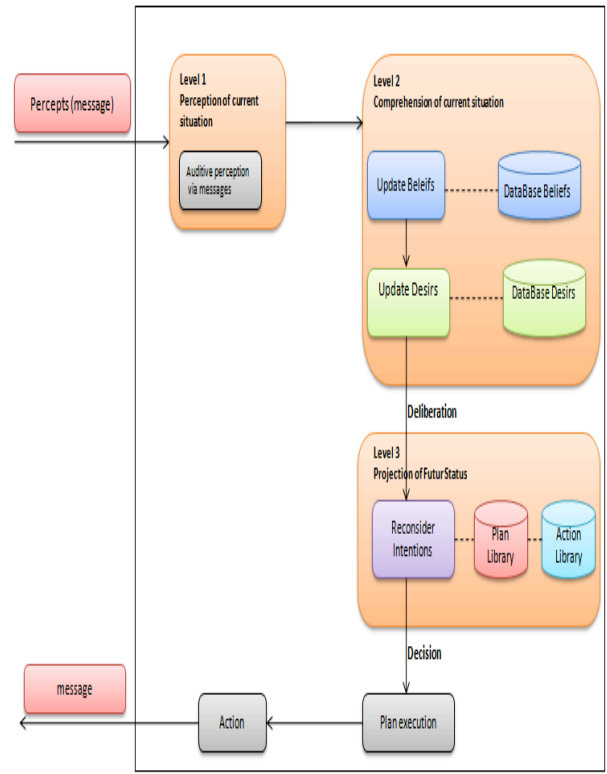

**Figure 4: Integration of situational awareness model [8] with the classic BDI model**

Referring to the above figure, messages concerning changes in the environment are received by an agent. These may be verbal messages from one agent to another, or from one agent to the team (broadcasted messages to all agents) and they take the form of logical messages, as shown in the right hand-column in figure 3 and detailed in [22]. The receipt of these messages constitutes the perception of the current situation (level 1 of SA). Agents have a database of things that they believe to be true in their world, e.g. the heart rate of the patient. Note that we speak of beliefs and not facts, since agents hold different beliefs of a situation; this allows agents' beliefs to differ, much as with humans in the real world. For example, one agent may believe that the patient is in a stable condition, while another does not. The BDI approach allows for beliefs to be different and even contradictory. This facility allows us to identify potential causes of misunderstandings within a group of agents. When new percepts are received, an agent's beliefs are accordingly updated and stored in the belief database. Agents also have goals that represent what they want to achieve. In multi-agent systems terminology, goals allow an agent to have pro-active (or goal directed behaviour). According to the agent's current goals and their current set of beliefs, they generate or update their desires. Desires represent a list of possible actions that an agent can undertaken in order to achieve their goals. An agent then deliberates which one of the desires to choose; this then becomes their intention (level 3 SA). In order to achieve an intention (which in effect is a goal or subgoal) an agent composes a plan, which is a list of actions to do. Actions may differ from agent to agent, e.g. only some members of the resuscitation may be allowed to administer medicine intravenously. Note that there is an existing library of plans available in order to reduce computational expense. An agent then executes the plan by working through the actions contained in the plan (e.g. perform tracheal intubation). The result of an action is a message. Following Searle and Austin's work on language [31] [5], which forms the basis for agent communication languages (ACLs) in multi-agent systems, a message is treated as an action that has an effect on the environment (i.e. it may be a percept that is received by other agents).

For further information of an actual BDI algorithm, a good explanation is given in [37].

Before discussing how this architecture has addressed the problems associated with the original SA model, we will apply the architecture to an example taken from our field studies. Imagine that the doctor asks the nurse, who is currently waiting for directives, to inject 1mg of adrenaline. She receives the message (level 1 SA) and according to this percept, her basic beliefs and desires are updated. In the basic beliefs a new belief will be added, "the patient needs an injection of adrenaline" and in the desires database a new desire: "inject adrenaline 1mg" is added (level 2 SA). Consequently the nurse will review her intentions, which previously was to wait for directives (injecting adrenaline has a higher priority than waiting). The new intention is to inject 1mg adrenaline. She will check if she can perform this action[1] and if so, she will send a message to the doctor that she agrees (level 3 SA).

## 6. Discussion and conclusion

A number of points have been addressed with this architecture. Firstly, we can see how communications have been encoded as logical messages (c.f. figure 3). We may also see how the simulator could be easily configured to experiment with the techniques covered in table 1. For example an agent can be made to send a message confirming the receipt and content of a request, thus simulating the 'check-back' techniques.

---

[1] Plan of actions to inject 1mg adrenaline: take the adrenaline ampoule, prepare a syringe of 1mg adrenaline, ask for confirmation from the physician to inject the adrenaline, wait for confirmation, inject. Note that this action plan already uses a communication protocol of asking and waiting for confirmation.

Secondly, there was a problem with the original SA model in that the cognitive processes of perceiving, understanding and projection are only described in general terms and are not operationalized [12]. We can see how the above architecture and the underlying BDI architecture makes these processes concrete, providing representations and allowing the trace of reasoning to be broken down and examined.

Concerning the issue that the levels in the original SA model are treated as sequential, we can see in the architecture how beliefs, desires and intensions may be constantly revised due to continuously incoming precepts. Therefore, for example, comprehension does not have to wait until all percepts are received.

We also saw that the original SA approach did not model the shared mental model of several people, but only described the internal model of the situation of a person.

The architecture above also only describes the situation as see from one person or agent, i.e. a mental model of the thought processes and elements in the surrounding environment. However, this is the architecture of each of the agents. If a macroscopic view of the agents is taken then it is possible to identify any overlap in the cognitive environments of a set of agents; i.e. the intersection formed by common beliefs and intentions. This allows us to identify the extent of the shared situation awareness amongst the agents, and importantly which of the agents does not hold beliefs that are held by other agents. Furthermore, by identifying contradictory beliefs it is possible to pinpoint the cause of misunderstandings or communication failures.

To conclude, this paper has addressed the issue of communication and mutual situation awareness in the case of a resuscitation team performing advanced cardiac life support.

Validation of our model has been undertaken at a preliminary level with data from field studies and with experts. However, a more in-depth validation is necessary if the model is to be used a firm basis for the simulator. Since we have adopted a strongly iterative approach to development, the current simulator requires further development and validation. Our architecture is based on the assumption of verbal communication. However, we are aware that non-verbal communications play an important role. In terms of the architecture this changes little since percepts can be visual as well as auditory. However, handling non-verbal communications in the simulator complicates the implementation. A 'can_hear' function has been implemented, drawing upon our previous work, but we would also need to implement a seeing behaviour (function) in the agents.